\documentclass[draftcls, onecolumn]{IEEEtran}

\usepackage{slashbox,colortbl,amsmath}
\usepackage[table]{xcolor}

\usepackage[applemac]{inputenc}
\usepackage[english]{babel}
\usepackage{graphicx}
\usepackage[]{subfigure}
\usepackage{color}
\usepackage{cite}
\usepackage{amssymb,amsmath,MnSymbol,wasysym,linearA,upgreek}
\newtheorem{theorem}{\textbf{Theorem}}
\newtheorem{lemma}[theorem]{\textbf{Lemma}}

\newtheorem{example}[theorem]{\textbf{Example}}

\newtheorem{proposition}[theorem]{\textbf{Proposition}}

\newtheorem{algo}[theorem]{\textbf{Algorithm}}

\newtheorem{defin}[theorem]{\textbf{Definition}}

\newcommand{\ba}{\begin{array}}
\newcommand{\ea}{\end{array}}

\newcommand{\nat}{\mathbb{N}}

\def\set#1#2{\{#1\mid\;#2\}}

\long\def\symbolfootnote[#1]#2{\begingroup%
\def\thefootnote{\fnsymbol{footnote}}\footnote[#1]{#2}\endgroup}
\begin{document}

\title{Testing experiments on synchronized Petri nets}
\author{Marco Pocci$^{(a)}$, Isabel Demongodin$^{(b)}$, Norbert Giambiasi$^{(c)}$, Alessandro Giua$^{(d)}$ \\
$^{(a)-(d)}$Laboratoire des Sciences de l'Information et des Systmes, University of Aix-Marseille, France\\
$^{(a),(d)}$Department of Electrical and Electronic Engineering, University of  Cagliari, Italy\\
\{$^{(a)}${marco.pocci}, $^{(b)}${isabel.demongodin}, $^{(c)}${norbert.giambiasi}\}@lsis.org, $^{(d)}${giua@diee.unica.it}}

\maketitle\thispagestyle{empty}

\begin{abstract}
Synchronizing sequences have been proposed in the late 60Õs to solve testing problems on systems modeled by finite state machines. Such sequences lead a system, seen as a black box, from an unknown current state to a known final one.

This paper presents a first investigation of the computation of synchronizing sequences for systems modeled by bounded synchronized Petri nets. In the first part of the paper, existing techniques for automata are adapted to this new setting. Later on, new approaches, that exploit the net structure to efficiently compute synchronizing sequences without an exhaustive enumeration of the state space, are
presented.
\end{abstract}

%


%
%

\section{Introduction}\label{introduction}
\IEEEPARstart{D}{ue} to increasingly larger size and rising complexity, the need of checking systems' performance increases and testing problems periodically resurface.


These problems have been introduced by the pioneering paper of Moore \cite{moore56}, where the main focus is to understand what can be inferred about the internal conditions of a system under test from external experiments. In his \emph{gedanken-experiment}, the system under investigation is a fixed semi-automaton seen as a black box. Lee and Yannakakis \cite{Lee-survey} have widely reviewed those problems and the techniques to solve them. They have stated five fundamental problems of testing: i) determining the final state after a test; ii) state identification; iii) state verification; iv) conformance testing; v) machine identification. Among these, the problem of determining the final state after a test is considered. This problem has been addressed and essentially completely solved using Mealy machines around 1960, using \emph{homing sequences} (HS) and \emph{synchronizing sequences} (SS).

The synchronization problem is the problem the reader deals with in this work. It concerns how to drive a system to a known state when its current state it is not known and when the outputs are not observable. This problem has many important applications and is of general interest.
It is of relevant importance for robotics and robotic manipulation \cite{Natarajan86,Natarajan89,ananichev2003}, when dealing with part handling and orienting problems in industrial automation such as part feeding, loading, assembly and packing. The reader can easily see that for example every device part, when arriving at manufacturing sites, needs to be sorted and oriented before assembly. Synchronization protocols have been developed to address global resource sharing in hierarchical real-time scheduling frameworks \cite{Heuvel99,Behnam_07}. An interesting automotive application can be found in \cite{Nolte09}.
Synchronization experiments have been done also in biocomputing, where Benenson \emph{et al.} \cite{Benenson01,Benenson03} have used DNA molecules as both software and hardware for finite automata of nano-scaling size. They have produced a solution of $3\times10^{12}$ identical automata  working in parallel. In order to synchronously bring each automaton to its "ready-to-restart" state, they have spiced it with a DNA molecule whose nucleotide sequence encodes a reset word. J\"urgensen \cite{Jurgensen2008} has surveyed synchronization issues from the point of view of coding theory in real life communication systems. He has presented the concept of synchronization in information channels, both in the absence  and in the presence of noise.
Synchronization is an important issue in network time protocol \cite{Gaderer10,Mills91}, where sharing of time information guaranties the correct internet system functioning.
Most of real systems, natural or man-built, have no integrated reset or cannot be equipped with. That is the case of digital circuits, where a reset circuit not only involves human intervention but increases the cost of the device itself reducing its effectiveness.
In this field Cho \emph{et al.} \cite{Cho93} have shown how to generate test cases for synchronous circuits with no reset.
When classic procedures fail due to large circuit size or because a synchronizing sequence does not exist, Lu \emph{et al.} \cite{Pomeranz96} propose a technique based on partial reset, i.e., special inputs that reset a subset of the flip-flops in the circuit leaving the other flip-flops at their current values. 
Hierons \cite{Hierons04} has presented a method to produce a test sequence with the minimum number of resets.
Nowadays the synchronizing theory is a field of very intensive research, motivated also by the famous \u{C}ern\' y conjecture \cite{cernia64}.
In 1964 J\'an \u{C}ern\' y has conjectured that $(n-1)^2$ is the upper bound for the length of the shortest SS for any $n-$state machine.
The conjecture is still open except for some special cases \cite{eppstein90},\cite{ananichev2003},\cite{trahtman90}.
Synchronization allows simple error recovery since, if an error is detected, a SS can be used to initialize the machine into a known state. That is why synchronization plays a key r\^ole in scientific contexts, without which all system behavior observations may become meaningless.
Thus the problem of determining which conditions admit a synchronization is an interesting challenge. This is the case of the \emph{road coloring problem}, where one is asked whether there exists a coloring, i.e., an edge labeling, such that the resulting automaton can be synchronized.
It was first stated by Adler in \cite{Adler77}. It has been investigated in various special cases and finally a positive solution has been presented by Trahtman in \cite{Trahtman09}, for which complexity analysis are provided \cite{Roman11}.

At present the problem of determining a synchronizing sequence has not  yet been investigated for Petri net (PN) models and only few works have addressed the broad area of testing in the PN framework.

The question of automatically testing PNs has been investigated by Jourdan and Bochmann in \cite{JourdanBochmann09}. They have adapted methods originally developed for \emph{Finite State Machines} (FSMs) and, classifying the possible occurring types of error, identified some cases where \emph{free choice} and \emph{$1$-safe} PNs \cite{Murata} provide more significant results especially in concurrent systems. Later the authors have extended their results also to \emph{$k$-safe} PNs \cite{Bochmann09}.
Zhu and He have given an interesting classification of testing criteria \cite{ZhuHe02} --- without testing algorithms --- and presented a theory of testing high-level Petri nets
by adapting some of their general results in testing concurrent software systems.

In the PN modeling framework, one of the main supervisory control tasks is to guide the system from a given initial marking to a desired one similarly to the synchronization problem. Yamalidou \emph{et al.} have presented a formulation based on linear optimization \cite{Yamalidou91,Yamalidou92}.
Giua \emph{et al.} have investigated the \emph{state estimation problem}, proposing an algorithm to calculate an estimate --- and a corresponding error bound --- for the actual marking of a given PN based on the observation of a word.
A different state estimation approach has been presented by Corona \emph{et al.} \cite{Giuaseatzu07}, for labelled PNs with silent transitions, i.e., transitions that do not produce any observation.
Similar techniques have been proposed by Lingxi \emph{et al.} in \cite{Lingxi09} to get a minimum estimate of initial markings, aiming to characterize the minimum number of resources required at the initialization for a variety of systems.

This paper is focused on bounded synchronized PNs and the SS problem here is first investigated.
The paper shows how the Mealy machine approach \cite{Lee-survey} can be easily adapted to systems represented by the class of bounded synchronized PNs.
Synchronized PNs, as introduced by Moalla \emph{et al.} in \cite{Moalla78}, are nets where a label associated with each transition corresponds to an external input event whose occurrence causes the firing of all marking enabled transitions having this label.
Note that SSs are independent of the output and this makes the synchronized PN a suitable model for such an analysis.
Then the authors consider a special class of Petri nets called \emph{state machines} (SMs) \cite{Murata}, characterized by the fact that each transition has a single input and a single output arc. Note that this model, albeit simple, is more general an automaton. In fact, while the reachability graph of a state machine with a single token is isomorphic --- assuming all places can be marked --- to the net itself, as the number of tokens $k$ in the net increases the reachability graph grows as $k^{m-1}$, where $m$ is the number of places in the net. It is shown that for strongly connected SM even in the case of multiple tokens, the existence of SS can be efficiently determined by just looking at the net structure, thus avoiding the state explosion problem.
These results are also extended to SMs that are not strongly connected {and to PN containing SM subnets. The effectiveness of the technique is proved via a toolbox we developed on Matlab \cite{Pocciweb}.}

The paper is organized as follows.
In Section~\ref{Background} the background on automata with inputs and PNs is provided.
Section~\ref{SS:MM} presents the classic SS construction method for automata with inputs.
Section~\ref{SS:PN} shows how to obtain SSs by adapting the classic method developed for automata with inputs to bounded synchronized PNs, via reachability graph construction.
Section~\ref{SMPN} proposes an original technique, based on path analysis, for efficiently determining SSs on strongly connected SMs.
Section~\ref{complexity} presents a short discussion of algorithm complexity.
The case of non-strongly connected SMs is investigated in Section~\ref{NSCPN}.
{In Section~\ref{sec:subSMs} our approaches are extended to nets containing state machine subnets. In Section~\ref{sec:ex_results} numerical results are presented, applying our tool to randomly generated SMs.
Finally, in Section~\ref{conclusion}, conclusions are drawn and open areas of research are outlined.}

\section{Background}\label{Background}
\subsection{Automata with inputs}\label{mm_formalisms}


An \emph{automaton with inputs} $\Lambda$ is a structure
$$\Lambda=(\chi, E, \delta),$$
 where $\chi$ and $E$ are finite and nonempty sets of states and input events respectively, and  $\delta: \chi\times E \rightarrow \chi$ is the state transition function.

When the automaton is in the current state $x\in \chi$ and receives an event $e \in E$, it reaches the next state specified by $\delta(x,e)$.

Note that $\delta$ is usually assumed to be a \emph{total function}, i.e., a function defined on each element $(x,e)$ of its domain. In such a case the automaton is called \emph{completely specified}.

The number of states and input events are respectively denoted by $n=|\chi|$, $p=|E|$. 
 One can extend the transition function $\delta$ 
from input events to sequences of input events as follows: a)
if $\varepsilon$ denotes the empty input sequence, $\delta(x,\varepsilon)=x$ for all $x\in \chi$; b) for all $e\in E$ and for all $w\in E^*$ it holds that $\delta(x,we)=\delta(\delta(x,w),e)$\footnote{Here $*$ denotes the Kleene star operator and $E^*$ represents the set of all sequences on alphabet $E$.}.

The transition function $\delta$ can also be extended
to a set of states as follows:
for a set of states $\chi'\subseteq \chi$, an input event $e\in E$ yields the set of states $ \chi''=\delta(\chi',e)=\bigcup_{x\in \chi'}\delta(x,e).$

A simple way to represent any automaton is a graph, where states and input events are respectively depicted as nodes and labelled arcs.


An automaton with inputs is said \emph{strongly connected} if there exists a directed path from any node of its graph to any other node.

The set of nodes of a non-strongly connected automaton can be partitioned into its maximal strongly connected components. A component is called
\emph{ergodic}, if its set of output arcs is included in its set of input arcs, \emph{transient}, otherwise.

An automaton contains at least one ergodic component and a strongly connected automaton consists of a single ergodic component.

%
%

\subsection{Place/Transition nets}

In this section, it is recalled the PN formalism used in the paper. For more details on PNs the reader is referred to \cite{Murata,BODavid04}.

A Petri net (PN), or more properly a \emph{Place/Transition net}, is a structure $$N=\left(P,T,Pre,Post\right),$$ where $P$ is the set of $m$ places, $T$ is the set of $q$ transitions, $Pre:\, P \times T \rightarrow \mathbb{N}$ and $Post:\, P \times T \rightarrow \mathbb{N}$ are the pre and post incidence functions that specify the weighted arcs.

A \emph{marking} is a vector $M:\, P \rightarrow \mathbb{N}$ that assigns to each place a nonnegative integer number of tokens; the marking of a place $p$ is denoted with $M(p)$. A marked PN is denoted $\langle N, M_0\rangle$.

A transition $t$ is enabled at $M$ iff $M \geq Pre(\cdot, t)$. An enabled transition may be fired yielding the marking $M^\prime=M + Post(\cdot, t) - Pre(\cdot, t)$. The set of enabled transitions at $M$ is denoted ${\cal E}(M)$.

$M[\sigma \rangle$ denotes that the sequence of transitions $\sigma=t_{1} \ldots t_{k}$  is enabled at $M$ and $M[\sigma \rangle M^\prime$ denotes that the firing of $\sigma$ from $M$ yields $M^\prime$.



A marking $M$ is said to be \emph{reachable} in $\langle N,
M_0\rangle$ iff there exists  a firing sequence $\sigma$
 such that $M_0[\sigma \rangle M$. The set of all markings reachable from $M_0$ defines the \emph{reachability set} of $\langle N, M_0\rangle$ and is denoted with $R(N,M_0)$.

\begin{figure}
 \centering
 \subfigure[]
   {\label{PN2}
   \includegraphics[scale=0.6]{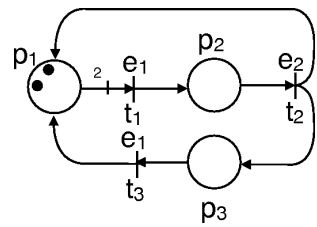}}\\
 \subfigure[]
   {\label{behavior2}
   \includegraphics[scale=0.7]{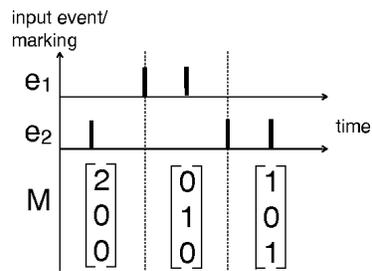}}
\caption{A synchronized PN (a) and a possible behavior (b).}
 \end{figure}

The $preset$ and $postset$ of a place $p$ are respectively denoted $^\bullet p$ and $p^\bullet$.
One can define the set of input transitions for a set of places $\hat P$ as the set $^\bullet \hat P=\{t: \forall p\in \hat P,\ t\in \ ^\bullet{p}\}$. Analogously the set of output transitions for a set of places $\hat P$ is the set $\hat P^\bullet =\{t: \forall p\in \hat P,\ t\in \ {p}^\bullet\}$.

%


\subsection{Synchronized Petri nets}
A \emph{synchronized PN} \cite{BODavid04} is a structure $\langle N, E,f \rangle$ such that: i) $N$ is a P/T net; ii) $E$ is an input alphabet of external events; iii) $f: T \to E$ is a labeling function that associates with each transition $t$ an input event $f(t)$.

Given an initial marking $M_0$, a marked \emph{synchronized PN} is a structure $\langle N, M_0, E, f\rangle$.

One extends the labeling function to sequences of transitions as follows: if $\sigma=t_1t_2\ldots t_k$ then $f^*(\sigma)=f(t_1)f(t_2)\ldots f(t_k)$.


The set $T_e$ of transitions associated with input event $e$ is defined as follows: $T_e= \set{t}{t\in T, f(t)=e}$. Equivalently all transitions in $T_e$ are said to be receptive to input event $e$.


The evolution of a synchronized PN is driven by input sequences as it follows. At marking $M$, transition $t \in T$ is fired iff:
\begin{enumerate}
  \item it is enabled, i.e., $t \in {\cal E}(M)$;
  \item the event $e = f(t)$ occurs.
\end{enumerate}
On the contrary, the occurrence of an event associated with a transition $t \not\in {\cal E}(M)$ does not produce any firing.
Note that a single server semantic is here adopted, i.e., when input event $e$ occurs, the enabled transitions in $T_e$ fire only once regardless of their enabling degree.

One writes $M\xrightarrow{w}M'$ to denote the fact that the application of input event sequence $w=e_{1} \ldots e_{k}$ from $M$ drives the net to $M^\prime$.

In Figure~\ref{PN2} is shown an example of synchronized PN. Note that labels next to each transition denote its name and the associated input event. In Figure~\ref{behavior2} the net evolution is presented over a possible input sequence $w=e_2e_1e_1e_2e_2$ starting from marking $M_0$.


%
%
%

In the rest of the paper, the reader will only deal with the class of bounded synchronized PNs that also satisfy the following structural restriction, that is common in the literature to ensure the determinism of the model:
\begin{equation}
\nexists p \;s.t. \qquad t,t'\in p^\bullet \; and \; f(t)=f(t').
\end{equation}

When an event occurs in a deterministic net, all enabled transitions receptive to that event can simultaneously fire.
Thus an input sequence $w=e_1 e_2 \cdots e_k \in E^*$ drives a deterministic net through the sequence of markings $M_0$, $M_1$, $M_2$, $\cdots$,  $M_k$ where $M_0$ is the initial marking and $$M_{i+1} =  M_i +  \sum_{t\in T_{e_{i+1}} \bigcap{\cal E}(M_i)} \left( Post(\cdot, t) - Pre(\cdot, t) \right).$$

\begin{example}
Consider the PN of Figure~\ref{PN2} and let $M=[2\, 0\, 1]^T$ be the current marking. Transitions $t_1$ and $t_3$ are enabled and upon the occurrence of event $e_1$ will simultaneously fire, yielding marking $M'=[1\, 1\, 0]^T$. Note that
markings $[0\, 1\, 1]^T$ and $[3\, 0\, 0]^T$, respectively obtained by the independent firing of $t_1$ and $t_3$, are never reachable.
\hfill $\blacksquare$
\end{example}


A marked PN $\langle N, M_0\rangle$ is said to be bounded if there exists a positive constant $k$ such that for all $M \in R(N,M_0)$, $M(p) \leq k\,\, \forall p\in P$. Such a net has a finite reachability set. In this case, the behavior of the net can be represented by the \emph{reachability graph} (RG), a directed graph whose vertices correspond to reachable markings and whose edges correspond to the transitions and the associated event causing a change of marking.

The graph in Figure~\ref{graph_specified} (disregarding the dashed edges) is the reachability graph of the PN in Figure~\ref{PN2}.

\subsection{State machine Petri nets}\label{subsec:SM}

Let first recall the definition of a state machine PN.

\begin{defin}[State machine PN]\em \cite{Murata}
A \emph{state machine} (SM) PN is an ordinary PN such that each transition $t$ has exactly one input place and exactly one output place, i.e.,
$$ \hspace{2,45cm}|^\bullet t|=|t^\bullet|=1 \qquad(\forall t\in T) \hspace{2,45cm} \blacksquare$$
\end{defin}

Observe that a SM $N=\left(P,T,Pre,Post\right)$ may also be represented by an \emph{associated  graph} ${\cal G}_N =(V,A)$ whose set of vertices $V=P$ coincides with set of places of the net, and whose set of arcs $A$ corresponds to the set of transitions of the net,  i.e.,
$ A \subseteq P \times P = \{ (p_i, p_j) \mid \exists t \in T, p_i = {^{\bullet}t}, p_j = t^{\bullet} \}.$

Such a graph can be partitioned into its maximal strongly connected components, analogously to the automata with inputs.
These components induce also a partition of the set of places of the corresponding SM.
\begin{defin}[Associated graph] \em
Given a SM $N=\left(P,T,Pre,Post\right)$, let ${\cal G}_N =(P,A)$ be its associated  graph. $P$ can be partitioned into \emph{components} as follows:
 $$P = P_1 \cupdot \cdots \cupdot P_k$$
 such that for all $i= 1, \ldots, k$ and $A_i = A \cap (P_i \times P_i)$ it holds that $(P_i,A_i)$  is a maximal strongly connected sub-graph of ${\cal G}_N$.
\hfill $\blacksquare$
\end{defin}

As discussed in Section~\ref{mm_formalisms}, components $P_1, \ldots P_k$ can be classified as transient or ergodic components.


\begin{defin}[Condensed graph]\label{def_cond_graph} \em
Given a SM $N=(P,T,$ $Pre,Post)$, its corresponding \emph{condensed graph} ${\cal C}(N)$ is defined as a graph where each node represents a maximal strongly connected component and whose edges represent the transitions connecting these components.
\hfill $\blacksquare$
\end{defin}

\begin{figure}[ht]
 \centering
 \subfigure[]
   {\label{PN2Er3Tr}
   \includegraphics[scale=0.6]{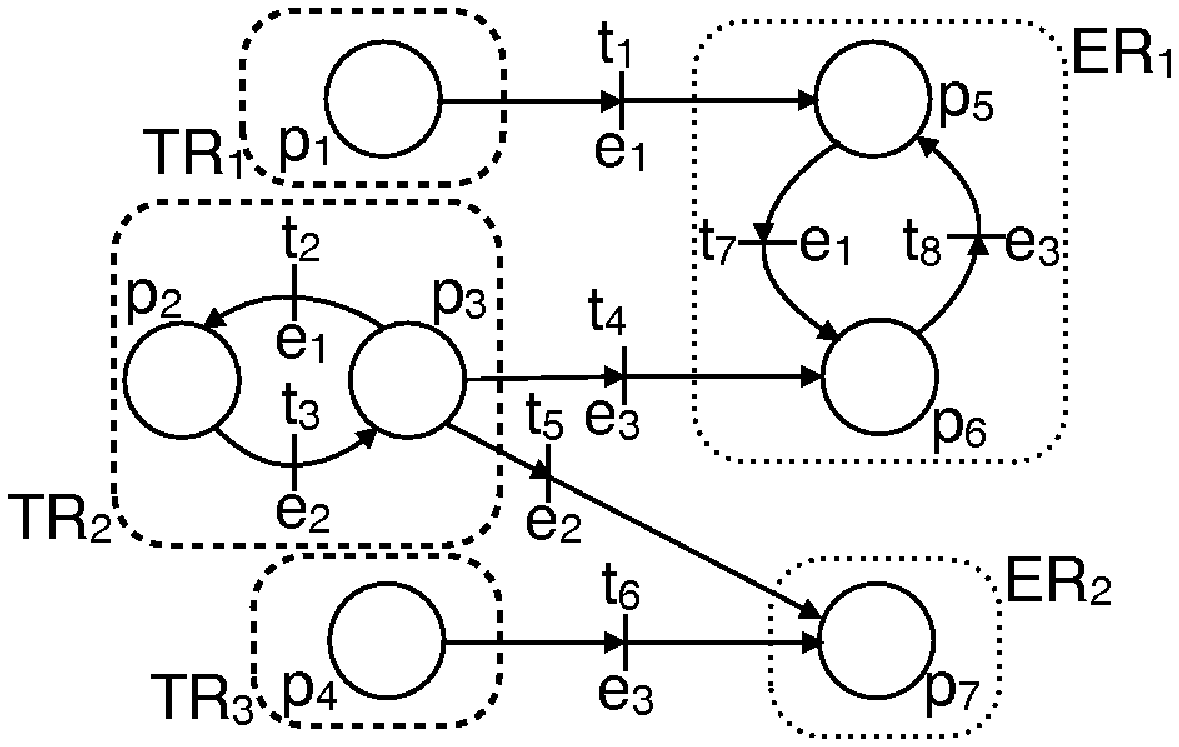}}\\
 \subfigure[]
   {\label{aux_graph}
   \includegraphics[scale=0.6]{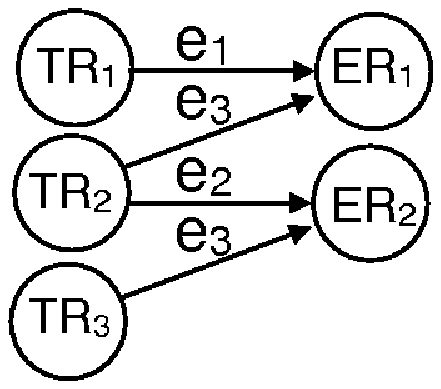}}
\caption{A not strongly connected PN (a) and its condensed graph (b).}
 \end{figure}

In Figure~\ref{PN2Er3Tr} it is shown an example of a synchronized SM which is not strongly connected. Transient and ergodic components are respectively identified by dashed and dotted boxes. For such a net the transient components are $TR_1=\{p_1\}$, $TR_2=\{p_2,\ p_3 \}$, $TR_3=\{p_4\}$ and the ergodic components are $ER_1=\{p_5,\ p_6\}$ and $ER_2=\{p_7\}$.
The corresponding ${\cal C}(N)$ is shown in Figure~\ref{aux_graph}, where subnets induced by each component are represented by single nodes.


\section{Synchronizing sequences for automata with inputs}\label{SS:MM}

In this section, the SS classic construction is presented by the aid of finite automata.

\begin{defin}[SSs on automata]\label{def_SS_automata} \em
Consider an automaton with inputs $\Lambda=(\chi, E, \delta)$ and a state $\bar{x}\in \chi$.
The input sequence $\bar{w}$ is called \emph{synchronizing for state $\bar{x}$} if it drives the automaton to $\bar{x}$, regardless of the initial state, i.e., $\forall x\in \chi$ it holds that $\delta(x,w)=\bar{x}$.
\hfill $\blacksquare$
\end{defin}


The information about the current state of $\Lambda$ after applying an input sequence $w$ is defined by the set $\phi(w)=\delta(\chi,w)$, called the \emph{current state uncertainty of $w$}.
In other words $w$ is a synchronizing sequence (SS) that takes the automaton to the final state $\bar{x}$ iff $\phi(w)=\{\bar{x}\}$.

The synchronizing tree method \cite{Hennie68,kohavi2} has been proposed to provide shortest SSs. Such a method is suitable only for small size systems, since the memory required to build up the tree is high, and becomes useless when the size grows. As a matter of fact the problem of finding shortest SSs is known to be NP-complete \cite{eppstein90}.

Two polynomial algorithms have been mainly used to provide SSs that are not necessarily the shortest. The so-called
\emph{greedy} and \emph{cycle} algorithms, respectively of Eppstein \cite{eppstein90} and Trahtman \cite{trahtman04}, that have equivalent complexity.

{The \emph{greedy} algorithm \cite{eppstein90}
 determines an input sequence that takes a given automaton, regardless of its initial state, to a known target state: note that the target state is determined by the algorithm and cannot be specified by the user.
Here we  propose a slightly different implementation of the greedy algorithm (see Algorithm~\ref{algo_ss}), that takes as input also a state $\bar{x}$ and determines a sequence that synchronizes to that state.

\begin{figure}[ht]
 \centering
 \subfigure[]
   {\label{graph_specified}
 \includegraphics[scale=0.7]{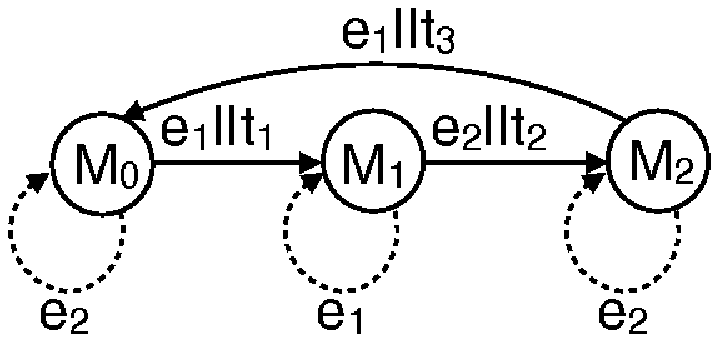}
 }\\
 \subfigure[]
   {\label{auxg2}
\includegraphics[scale=0.7]{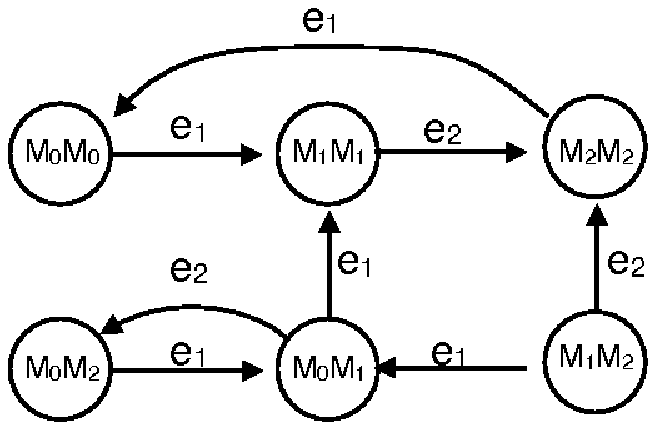}}
\caption{The completely specified RG (a) and the corresponding auxiliary graph (b) of the PN in Figure~\ref{PN2}.}
 \end{figure}

This algorithm is later used as a building block to determine a SS to reach a given marking among those in the reachability set of a bounded PN.}

\begin{defin}[Auxiliary graph]\label{def:aux_graph}
\em
Given an automaton with inputs $\Lambda$ with $n$ states, let ${\cal A}(\Lambda)$ be its \emph{auxiliary graph}. ${\cal A}(\Lambda)$ contains $n(n+1)/2$ nodes, one for every unordered pair $(x', x'')$ of states of $\Lambda$, including pairs $(x, x)$ of identical states. There exists an edge from node $(x', x'')$ to $(\hat{x}', \hat{x}'')$ labeled with an input event $e\in E$ iff $\delta(x',e)=\hat{x}'$ and $\delta(x'',e)=\hat{x}''$.
\hspace{2,1cm}\hfill $\blacksquare$
\end{defin}


\begin{algo}(Greedy computation of SSs on automata with inputs)\label{algo_ss}
\em
\\\textbf{Input:} An auxiliary graph ${\cal A}(\Lambda)$, associated with an automaton with inputs $\Lambda=(\chi, E, \delta)$, and a target state $\bar{x}\in \chi$.
\\\textbf{Ouput:} A SS $\bar{w}$ for state $\bar{x}$.

\begin{enumerate}
\item[{\bf 1.}] Let $i=0$.
\item[{\bf 2.}] Let $w_0= \varepsilon$, the empty initial input sequence.
\item[{\bf 3.}] Let $\phi(w_0)=\chi$, the initial current state uncertainty.
\item[{\bf 4.}] While $\phi(w_{i})\neq \{\bar{x}\}$, do
\begin{enumerate}
\item[{\bf 4.1.}] $i=i+1$.
\item[{\bf 4.2.}] Pick two states $x, x'\in \phi(w_{i-1})$ such that $x\neq x'$.

\item[{\bf 4.3.}] {\bf If}  there does not exist any path in ${\cal A}(\Lambda)$ from node $(x', x'')$ to $(\bar{x}, \bar{x})$, stop the computation, there exists no SS for $\bar{x}$.

\item[] {\bf Else} find the shortest path from node $(x', x'')$ to $(\bar{x}, \bar{x})$ and let $w$ be the input sequence along this path, do
\begin{enumerate}
\item[]{\bf 4.3.1.} $w_{i}=w_{i-1}w$
\item[]{\bf 4.3.2.} $\phi(w_i)=\delta(\phi(w_{i-1}),w)$.
\end{enumerate}
\end{enumerate}
\item[{\bf 5.}] $\bar{w}=w_{i}$.\hfill $\blacksquare$
\end{enumerate}
\end{algo}

The following theorem provides a necessary and sufficient condition for the existence of a SS for a target final state.


\begin{theorem} \label{teo:reach_cond}\em
The following three propositions are equivalent.
\begin{enumerate}
\item Given an automaton with inputs $\Lambda=(\chi, E, \delta)$, there exists a SS for state $\bar{x}\in \chi$;
\item ${\cal A}(\Lambda)$ contains a path from every node $(x',x'')$, where $x',x''\in \chi$, to node $(\bar{x}, \bar{x})$;
\item Algorithm~\ref{algo_ss} determines a SS $\bar{w}$ for state $\bar{x}\in \chi$ at step~5., if there exists any SS.
\end{enumerate}
\end{theorem}

\begin{table}[ht]
\begin{center}
\begin{tabular}{|c|c|}
\hline $M_0$   &  $[2\, 0\, 0]^T$  \\
\hline $M_1$   &  $[0\, 1\, 0]^T$  \\
\hline $M_2$   &  $[1\, 0\, 1]^T$  \\
\hline
\end{tabular}
\vspace{0.1cm}
\end{center}
\caption{Markings of the PN in Figure~\ref{PN2}.}
\label{table_M_G1}
\end{table}

\IEEEproof {[1) implies 2)] If there exists a SS for state $\bar{x}\in \chi$, there exists an input sequence ${w}$ for $\bar{x}$ s.t. for any $x',x''\in \chi$ it holds that $\delta(x',{w})=\delta(x'',{w})=\bar{x}$. Hence there exists a path labeled ${w}$ from any $(x',x'')$ to $(\bar{x},\bar{x})$.

{[2) implies 3)]} Consider iteration $i$ of the while loop of Algorithm~\ref{algo_ss}. If there exists a path labeled $w$ from any $(x',x'')$ to $(\bar{x},\bar{x})$, then it holds that $\delta(\{x',x''\},w)=\{\bar{x}\}$. Hence the following inequality holds:
$$|\phi(w_i)| = |\delta(\phi(w_{i-1})\backslash\{x,x'\},w)\cup\{\bar{x}\}| \leq  |\phi(w_i)|-1.$$

The existence of such a sequence for every couple of states $x',x'' \in \chi$ assures that the current state uncertainty will be reduced to singleton $\{\bar{x}\}$ after no more than $n$ iteration.

{[3) implies 1)]} Since Algorithm~\ref{algo_ss} requires the current state uncertainty to be singleton and uses it as a stop criterium, if it terminates at step~5., then the sequence found is clearly a SS. \hfill $\square$}


%

One can easily understand that, when the automaton is not strongly connected, the above reachability condition will be verified only when there exists only one ergodic component and there may exist a SS only for those states belonging to this ergodic component.


%
%


\section{Synchronizing sequences for bounded synchronized PNs}\label{SS:PN}

When computing a SS for real systems modeled by automata, it is assumed that a complete description of the model in terms of space-set, input events and transition function is given. The idea is that the test generator knows all possible states in which the system may be.


A similar notion can be given
for Petri nets, where equivalently one can say that the test generator knows a "starting state", i.e., a possible state, of the system and the initial uncertainty coincides with the set of states reachable from this starting state.

In a synchronization problem via PNs, it is given a Petri net $N$ and a \emph{starting marking} $M_0$. 
The current marking $M$ is unknown, but it is assumed to be reachable from $M_0$.


This starting marking, together with the firing rules, provides a characterization of the initial state uncertainty, given by ${\cal M}_0 = R(N,M_0)$.
The goal is to find an input sequence that, { regardless of the initial marking}, drives the net to a known marking $\bar{M}\in R(N,M_0)$.





Given a synchronized PN $\langle N,E,f \rangle$, a straightforward approach to determine a SS consists in adapting the existing approach for automata to the reachability graph (RG).

It is easy to verify that this direct adaptation presents one shortcoming that makes it not always applicable: the greedy approach requires the graph to be completely specified, while in a RG of a PN this condition is not always true. In fact, from a marking not all transitions are necessarily enabled, causing the RG of the PN to be partially specified.
In order to use the aforementioned approach it is necessary to turn its RG ${\cal G}$ into a completely specified $\tilde{\cal G}$.

\begin{example}
Consider the PN in Figure~\ref{PN2}. The current marking $M=[2\ 0\ 0]^T$ enables only transition $t_1$, then all events not associated with $t_1$ are not specified. Hence for that marking one adds a self loop labelled $e_2$ and so on for the rest of the reachable markings.
 \hfill $\blacksquare$
\end{example}

In Figure~\ref{graph_specified} is shown the RG of the PN in Figure~\ref{PN2}. Note that dashed edges are added in order to make it completely specified.
In Figure~\ref{auxg2} is shown the corresponding auxiliary graph of the RG in Figure~\ref{graph_specified}.


One can summarize the modified approach for PNs in the following algorithm.
\begin{algo}(RG computation of SSs on synchronized PNs)
 \label{algo_ss_PN}\em
 \\\textbf{Input:} A bounded synchronized PN $\langle N,E,f\rangle$, a starting marking $M_0$ and a target marking $\bar{M}$.
\\\textbf{Ouput:} A SS $\bar{w}$ for marking $\bar{M}\in R(N,M_0)$.
\em

\begin{enumerate}
  \item[{\bf 1.}] Let ${\cal G}$ be the reachability graph of $\langle N,M_0 \rangle$.
  \item[{\bf 2.}] Let $\tilde{\cal G}$ be the modified reachability  graph obtained by completing ${\cal G}$, then by adding a self loop labelled $e$, i.e., $\forall M \in {\cal G}$ and $\forall e\in E$ s.t. $\nexists t\in T_e \cap {\cal E}(M)$.
  \item[{\bf 3.}] Construct the corresponding auxiliary graph ${\cal A}(\tilde{\cal G})$.
  \item[{\bf 4.}] A SS for marking $\bar{M}$, if such a sequence exists,
  is given by the direct application of Algorithm~\ref{algo_ss} to ${\cal A}(\tilde{\cal G})$, having $\bar{M}$ as target.\hfill $\blacksquare$
\end{enumerate}

\end{algo}

The following proposition can now be stated.

\begin{proposition}\label{teo:reach_condPN}\em
Given a bounded synchronized PN $N$ and a starting marking $M_0$, there exists a SS leading to a marking $\bar{M} \in R(N,M_0)$ iff the reachability condition on its auxiliary graph ${\cal A}(\tilde{\cal G})$ is verified, i.e., there is a path from every node $(M_i,M_j)$, with $M_i,M_j \in R(N,M_0)$, to node $(\bar{M}, \bar{M})$.
\end{proposition}

\IEEEproof{Consider a marked PN net $\langle N,M_0\rangle$ and its RG ${\cal G}$. Given a marking $M\in R(N,M_0)$, 
a sequence $\sigma=t_{j1}t_{j2}\ldots t_{jp}$ generates the trajectory $M[t_{j1}\rangle M_1[t_{j2}\ldots t_{jp}\rangle M_p$ iff there exists an oriented path $\gamma=M t_{j1} M_1 t_{j2}\ldots t_{jp} M_p$ in ${\cal G}$.
The same equivalence holds between a synchronized PN and its completely specified RG $\tilde{ \cal G}$. Thus an input sequence $w=e_{j1}e_{j2}\ldots e_{jp}$ drives the net from $M$ to $M_p$ iff there exists an oriented path $\gamma=M e_{j1} M_1 e_{j2}\ldots e_{jp} M_p$ in $\tilde{\cal G}$.

Since the completely specified RG $\tilde{\cal G}$ can be considered as an automaton whose behavior is equivalent to that of the synchronized PN, one can obtain a SS via Algorithm~\ref{algo_ss}.\hfill $\square$}

\section{Synchronizing sequences on strongly connected state machines}\label{SMPN}

Consider a strongly connected SM defined in Section~\ref{subsec:SM}. Knowing the number of tokens $k$ initially contained in the net --- regardless of their initial distribution --- is sufficient to exactly determine the reachability set of the net: in fact, the number of tokens will remain constant as the net evolves and any distribution of the $k$ tokens can be reached.

If a SM is not strongly connected, knowing the number of tokens $k$ initially contained in the net --- but not their initial distribution
--- will give a larger approximation of the reachability set that may be used to design a SS. The knowledge of the number of tokens initially contained in each component --- but not their initial distribution within each component --- will provide an exact characterization of the reachability set.

This new setting aims to determine a SS without constructing the whole state-space. Hence a new formal definition of SS for SMs has to be given.

\begin{defin}[SS on state machine PNs]\label{def_SS1} \em
Given a  synchronized SM $\langle N, E,f \rangle$, assume that the initial marking $M_0$ is not given but is known to belong to a set
$${\cal M}_0 = \{ M \in \nat^m \mid \sum_i M(p_i) = k \}.$$
$\bar{w}$ is called a $k$-SS if for all $M \in {\cal M}_0$ it holds $M\xrightarrow{\bar{w}}\bar{M}$.
\hfill $\blacksquare$
\end{defin}

In this section, we first analyze the problem of determining a $1$-SS and then address the more general $k$-SS, starting from $1$-SS.

\subsection{$1$-SS on strongly connected state machines}

In this subsection we present a particular technique to determine $1$-SSs via sufficient conditions over the net structure.
Such a technique can be more efficient than the approach presented in Algorithm~\ref{algo_ss_PN}, as discussed later in Section~\ref{complexity}.

Let us first give the definition of directed path.

\begin{defin}[Directed path]\label{path} \em
Given a  SM $N=(P,T,$ $Pre,Post)$, an alternated sequence of places and transitions $\rho= \langle p'_{0} t'_{1} p'_{1} t'_{2} \cdots t'_{r} p'_{r} \rangle$ is called a \emph{directed path} if $\forall i=0,\ldots r$ and $\forall j=1,\ldots r$ it holds: i) $p'_i \in P$ and $t'_j\in T$; ii) $p'_i \in {^{\bullet} t'_{j+1}}$ and $t'_j\in {^{\bullet} p'_{i}}$. A path non-containing any repeated place is called \emph{elementary}.
\hfill $\blacksquare$
\end{defin}

The notion of \emph{synchronizing transition sequence} for a set of places $\hat P$ and a specific place $\bar{p}\in \hat P$ can now be given.

\begin{defin}(Synchronizing transition sequence)\label{synch_path} \em Given a synchronized SM $\langle N, E,f \rangle$, let  $ \rho(\hat P, \bar p) =\langle p'_{0} t'_{1} p'_{1} t'_{2} \cdots  t'_{r} p'_{r} \rangle$, with $\bar{p} = p'_{r}$, be a directed path in $N=(P,T,Pre,Post)$ that visits all places in $\hat P \subseteq P$ and ends in $\bar{p} \in \hat P$, with $\bar p \neq p_j$ for $j=0, 1, \ldots, r-1$. Let $\sigma$ be the firing sequence obtained by removing all places from $\rho(\hat P, \bar{p})$. Such a sequence is called a \emph{synchronizing transition sequence} for $\hat P$ and $\bar{p}$ if
\begin{itemize}
\item [C1)] $\nexists t,t' \in \hat P^{\bullet}$ such that  $t \in \sigma$, $t'\not \in \sigma,$ and $f(t)=f(t')$.
\item [C2)] $\forall p'_i,p'_k\in \rho(\hat P, \bar{p}): $ if $ p'_i=p'_k$ and $i<k$ it holds that $f(t'_j) \neq f(t'_{k})$ for $j =1,\ldots k-1$.\hfill $\blacksquare$
\end{itemize}
\end{defin}

In simple words, condition C1) requires that there is no transitions exiting $\hat{P}$ and sharing the same label of a transition in $\sigma$.
Condition C2) requires that if a place is visited multiple times, its ingoing transition does not share the same label with any of the transitions in the path.

\begin{figure}
\centering
  \includegraphics[scale=0.6]{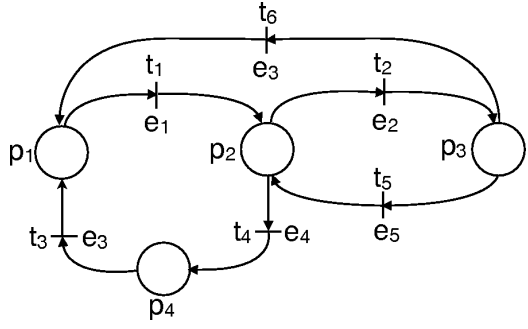}
  \caption{A strongly connected synchronized SM}\label{FIG:1token}
\end{figure}

A first result related to the existence of a SS for SMs with a single token can now be stated.

\begin{proposition}\label{propo_1token}\em
Consider a strongly connected synchronized SM $\langle N, E,f \rangle$ containing a single token. Let $\sigma$ be a synchronizing transition sequence for $P$ and $\bar{p}\in P$. Then $\bar{w}=f^*(\sigma)$ is a $1$-SS for marking $\bar{M}$ that assigns the token to place $\bar{p}$, i.e.,
$$\begin{array}{ccc} \bar{M}:\;\bar{M}(p) & = &
\left\{   \begin{array}{lll}
     1 \;\;if\;\; p=\bar{p}, \\
     0 \;\;otherwise. \\
   \end{array} \right.\end{array}$$
\end{proposition}
\IEEEproof{
Let $\sigma= t'_1 \cdots  t'_{r}$ be the synchronizing transition sequence found and $\rho(P,\bar{p}) = \langle p'_{0} t'_{1} p'_{1} t'_{2} \cdots t'_{r} p'_{r}\rangle$, with $\bar{p} = p'_{r}$, be the corresponding path (not necessarily elementary).

Let $\bar{w}$ be the corresponding input event sequence, i.e., $\bar{w}=f^*(\sigma)= e'_{0} \cdots e'_{r}$.

We first prove that after the occurrence of event $e_{1}$ the token can only be in a place $p'_{k}$ such that $k \geq 1$. Assume, in fact,  the token is initially in place $p'_i$ and event $e'_1$ occurs. Two different cases have to be treated.
If  $i=0$, then by definition of $\sigma$, the token is certainly driven to place $p'_1$. If  $i\geq 1 $, two further sub-cases are possible: a) no output transition of $p'_i$ has label $e_1$, i.e.,  $T_{e_1} \cap {p'_i}^\bullet = \emptyset$, and the token will not not move; b) an output transition of $p'_i$ has label $e_1$ and its firing moves the token to some place $p'_j$, with $j>i$.

The last result follows from conditions C1) and C2) of Definition~\ref{synch_path}. In fact,  condition C1) assures that only transitions belonging to $\sigma$ are receptive to $e_{1}$; thus the token can only be driven along the chosen path. Besides condition C2) assures that the token cannot go back in the upstream path along the sequence.

By repeating this argument, we can show that after the application of event $e_{i}$, for $i=2, \ldots, r$, the token can only be in a place $p'_{k}$ such that $k \geq i$, hence this ensures that when all events in the input sequence $\bar{w}$ have been applied the token will be in place $\bar p$. \hfill $\square$}

{Next algorithm shows how Proposition~\ref{propo_1token} can be effectively used to compute a $1$-SS.
In this, function $\tau: P\times T \rightarrow T$ (resp. $\pi: P\times T \rightarrow P$) returns the set of transitions (resp. places) visited by directed path $\rho$. Function $start: P\times T \rightarrow P$ determines the last place has been added to $\rho$. For instance, consider the path $\rho= \langle p_{r} t_{r} p_{r-1} t_{r-1} \cdots t_{1} p_{0} \rangle$. It holds that $\tau(\rho)=\{t_{1}, t_{2}, \cdots , t_{r}\}$,
$\pi(\rho)=\{p_{0}, p_{1}, \ldots ,  p_{r}\}$ and $start(\rho)=p_{r}$.

\begin{algo}(STS computation of $1$-SSs on synchronized PNs)
 \label{algo_ss_STS}\em
 \\\textbf{Input:} A SM PN $N=\left(P,T,Pre,Post\right)$ and a target place $\bar{p}$.
\\\textbf{Ouput:} a $1$-SS $w$ that drives the token to place $\bar{p}$.
\em
\begin{enumerate}
	\item[{\bf 1.}] $\rho= \bar{p}$, ${\cal R}=\{\rho\}$;
	\item[{\bf 2.}] $flag \coloneq false$;
	\item[{\bf 3.}] \textbf{while} $flag=false\;\vee \;{\cal R}\neq \emptyset$
	\begin{enumerate}
		\item[{\bf a.}] pick $\rho\in{\cal R}:\,|\rho|={\displaystyle\max_{{\rho'\in{\cal R}}} |\rho'|}$;
		
		\item[{\bf b.}] $p\coloneq start(\rho)$;
		\item[{\bf c.}] ${\cal T}\coloneq{^\bullet{p}} \backslash (\tau(\rho)\cup \bar{p}^\bullet)$;
		\item[{\bf d.}] \textbf{while} ${flag=false}\;\vee \;{\cal T}\neq \emptyset$,
			\begin{enumerate}
			\item [{\bf i.}] pick $t\in {\cal T}$, $\rho'\coloneq{^\bullet{t}} t \rho$;
			
			\item [{\bf ii.}] \textbf{if} $\rho'$ does not satisfiy C2), {\bf then goto} step {\bf 3.d.v.}
			\item [] \textbf{end if}
			
			\item [{\bf iii.}] \textbf{if} $\pi(\rho)=P$,
				\begin{enumerate}
				\item [$-$] \textbf{if} $\rho'$ satisfies C1),  \textbf{then} $flag\coloneq true						$.
				\item [] {\bf else}, {\bf goto} step {\bf 3.d.v.}
				\item [] {\bf end if}
				\end{enumerate}
			\item []\textbf{end if}
			
			\item[{\bf iv.}]  ${\cal R}\coloneq{\cal R}\cup\{\rho'\}$
			\item [{\bf v.}] ${\cal T}\coloneq{\cal T}\backslash \{t\}$;

			\end{enumerate}
		\item[] \textbf{end while}
		\item[\textbf{e.}] ${\cal R}\coloneq{\cal R}\backslash \{\rho\}$.
	\end{enumerate}
	\item[] \textbf{end while}
\item[{\bf 4.}] {\bf if} $flag=false$,
	\begin{enumerate}
	\item[{\bf a}] {\bf then} no STS exists;
	\end{enumerate}
\item[] {\bf else}
	\begin{enumerate}
	\item[{\bf b}]  pick $\rho\in{\cal R}:\,|\rho|={\displaystyle\max_{{\rho'\in{\cal R}}} |\rho'|}$;
	\item[{\bf c}]  let $\sigma$ be the firing sequence obtained removing all places from $\rho$;
	\item[{\bf d}]  $w\coloneq f^*(\sigma)$.
	\end{enumerate}
\item[] {\bf end if}\hfill $\blacksquare$
\end{enumerate}
\end{algo}

The algorithm 
computes a synchronizing transition sequence. It starts from desired place $\bar{p}$ (step~{\bf 1.}) and puts the path of zero length $\rho= \bar{p}$ into ${\cal R}$, which contains the set of path to be analyzed. The net is explored using a backward search until either a STS has been found, i.e.,  the $flag$ is true, or there are no more paths to analyze, i.e., ${\cal R} \neq\emptyset$ (step~{\bf 3.}).

Once a path $\rho$ is selected, we consider the set of transitions ${\cal T}$ inputting its start place $p$ that i) have not already been visited in the path; ii) do not output from the final place $\bar{p}$ (step~{\bf 3.c.}).

For all new paths $\rho'$, obtained adding to $\rho$ one transition in ${\cal T}$ and its input place (step~{\bf 3.d.i.}), we do the following. First, we check condition C2) (step {\bf 3.d.ii}), which must hold for all prefixes of the final path. If it does not hold, we discard the path going to step {\bf 3.d.v.}. Then we check if  $\rho'$ contains all places: in this case, if it satisfies condition C1) (step {\bf 3.d.iii}) we stop the algorithm (flag=true), 
else we discard it, 
going to step {\bf 3.d.v.}.
All new not discarded paths, are added to set ${\cal R}$ to be later explored (step {\bf 3.d.iv}).
 

When all transitions in ${\cal T}$ have been evaluated, path $\rho$ is then removed from ${\cal R}$ (step~{\bf 3.e.}).

In step~{\bf 4.}, if the flag is set to false, there is no $1$-SS constructible via the STS approach. Otherwise the path of maximum length is contained in ${\cal R}$ and it defines an STS. 

Paths are constructed via a depth-first-search, as ensured by the condition of step~{\bf 3.a.} that always picks (one of) the longest path(s). We could implement a \emph{breadth-first-search} by picking --- at the same step --- the shortest $\rho\in {\cal R}$, to ensure for the shortest STS solution if found.
}

\begin{example}\label{EX:1token}
Consider the strongly connected SM in Figure~\ref{FIG:1token}. The objective is to find a $1$-SS that leads the system to the marking $[ 0\, 0\, 0\, 1 ]^T$. Let $\rho=\langle p_1t_1p_2t_2p_3t_5p_2t_4p_4\rangle$ be the directed path that contains all the places, ending in $p_4$, and $\sigma= t_1t_2t_5t_4$ the synchronizing transition sequence for $P$ and place $p_4$. Sequence $\bar{w}=f^*(\sigma)=e_1e_2e_5e_4$ is a $1$-SS.
\hfill $\blacksquare$
\end{example}

Note that condition C1) of Definition~\ref{synch_path} is sufficient to assure the sequence to be a synchronizing one if $\rho$ is an elementary path.

The conditions given by Proposition~\ref{propo_1token} for the existence of a $1$-SS are sufficient but not necessary.

Although one determines a $1$-SS by just analyzing the net structure --- avoiding then the RG and the auxiliary graph construction and consistently reducing the complexity ---, the conditions required are very restrictive.

In fact there are SMs for which those conditions do not hold but that still have a $1$-SS.

\begin{example}\label{EX:1token_bis}
Consider again the strongly connected SM in Figure~\ref{FIG:1token} with one token and suppose $f(t_5)=f(t_2)=e_2$. This time one aims to find a $1$-SS that leads the system to the marking $[ 1\, 0\, 0\, 0 ]^T$.
There clearly exists no synchronizing transition sequence with such a change of the labeling function, hence no $1$-SS can be determined by Proposition~\ref{propo_1token}. Despite this, one easily finds the $1$-SS $\bar{w}=e_4e_3$ by the way of Algorithm~\ref{algo_ss_PN}.
\hfill $\blacksquare$
\end{example}

Note that, when conditions required by Proposition~\ref{propo_1token} do not hold, one can always determine a SS using Algorithm~\ref{algo_ss_PN}, obviously with an increased complexity as shown later in Section~\ref{complexity}.

\subsection{$k$-SS on strongly connected state machines}
We now consider the problem of determining a $k$-SS for nets with k tokens.

%
%



\begin{proposition}\label{propo_ktokenSM}\em
Consider a strongly connected synchronized SM $\langle N, E,f\rangle$ containing $k$ tokens.

Let $\sigma$ be a synchronizing transition sequence for $P$ and $\bar{p}\in P$ and $w=f^*(\sigma)$ a $1$-SS.
$w^k$ is a $k$-SS that moves all $k$ tokens to place $\bar{p}$, such that:

$$\begin{array}{ccc} \bar{M}:\;\bar{M}(p) & = &
\left\{   \begin{array}{lll}
     k \;\;if\;\; p=\bar{p}, \\
     0 \;\;otherwise. \\
   \end{array} \right.\end{array}$$\end{proposition}
\IEEEproof{
Consider a first application of $w$, at least one token is driven to $\bar{p}$.
Because of condition C2) and of the fact that the directed path does not pass through $\bar{p}$, none of the output transitions of this place is receptive to some event in $w$.
Hence every application of $w$ does not move the token from $\bar{p}$ and takes the $k$ tokens at least one by one to place $\bar{p}$.
\hfill $\square$}

\begin{example}\label{EX:ktoken}\rm
Consider the SM of Example~\ref{EX:1token}, where $w=e_1e_2e_5e_4$ is the $1$-SS previously found. Let the PN have 2 tokens.
It holds that $w^2=e_1e_2e_5e_4e_1e_2e_5e_4$ is a 2-SS, leading the net to the desired final marking $\bar{M}=[0\,0\,0\,2]^T$.
\hfill $\blacksquare$
\end{example}

The previous propositions show that determining a synchronizing transition sequence allows to readily construct not only a $1$-SS but a $k$-SS for an arbitrary $k$. However not all SSs can be obtained in this way.

Thus we consider the following problem: given any arbitrary $1$-SS, constructed not from a synchronizing transition sequence but by using
 Algorithm~\ref{algo_ss_PN}, does Proposition \ref{propo_ktokenSM} apply so that we can use it to construct a $k$-SS?
 Unfortunately this is not the case, as shown by the next example.

\begin{example}\label{EX:ktoken_false}\rm
Consider the SM of Example~\ref{EX:1token} and let $w=e_3e_1e_2e_5e_4$ be a $1$-SS for $p_4$. Let the PN have 2 tokens.
It is easy to see that $w^2$ is not a 2-SS, since only one token out of two would be driven to $p_4$.
\hfill $\blacksquare$
\end{example}

It is however possible to provide a sufficient condition for an arbitrary $1$-SS to ensure that, concatenating it $k$ times, a $k$-SS is obtained.

\begin{proposition}\label{propo_ktoken}\em
Consider a strongly connected synchronized SM $\langle N, E,f\rangle$ containing $k$ tokens.
Let $w$ be a $1$-SS for a target marking $\bar{M}$, such that $\bar{M}(p)= 1$ if $p=\bar{p}$, otherwise  $\bar{M}(p)= 0$.

If for all $t\in \bar{p}\,^\bullet$ it holds that $f(t) \not \in w$, i.e., sequence $w$ does not contain any symbol labeling an output transition of place $\bar{p}$, then $w^k$ is a $k$-SS  for a target marking $\bar{M}_k$, such that $\bar{M}_k(p)= k$ if $p=\bar{p}$, otherwise  $\bar{M}_k(p)= 0$.
\end{proposition}

\IEEEproof{
 During the first application of $w$, at least one of the tokens is driven to $\bar{p}$. Any further application of $w$ moves to $\bar{p}$ at least one of the tokens not in this place, and does not move the tokens already in $ \bar{p}$, as none of its output transitions is receptive to any event in $w$. Thus $w^k$ takes the $k$ tokens to place $\bar{p}$.
\hfill $\square$}

%
%
%
%
%

\section{A discussion on computational complexity}\label{complexity}

In this section we give an estimate of the computational complexity of both RG and STS approaches for $1$-SS construction.

\subsection{Complexity via reachability graph analysis}

The \emph{greedy} and the \emph{cycle} algorithms work both in $O(n^3+|E|n^2)$ time, where $n$ and $|E|$ are, resp.,  the number of states and the input alphabet cardinality of the  automaton. Proofs can be respectively found in \cite{eppstein90} and \cite{trahtman04}.

Such algorithms are applicable to synchronized PNs by first exhaustively enumerating the state space of the net, i.e., constructing its RG.
Although alternative techniques are proposed to decrease its complexity (e.g. \cite{Buchholz97,Ahmad09}), the RG generation suffers from the problem of exponential space and time complexity. In particular, for a SM the reachability set of markings can significantly increase with the number of tokens under the following expression.

\begin{theorem} \label{teo:spaceSM}\rm
Given a strongly connected SM $N=(P,T, Pre,Post)$, with
$k$  tokens, let $m$ be the number of its places.
The RG ${\cal G}$ of this net has a number of nodes equal to:
$$\binom{m+k-1}{m-1} \leq \frac{1}{(m-1)!}k^{m-1}$$
\end{theorem}
\IEEEproof{Consider the given net once a new node is added.
It can be easily shown that the reachability set cardinality is given by the following formula:

$$\begin{array}{l}
\displaystyle   |{\cal G}(m+1,k)|=\sum^k_{i=0}|{\cal G}(m,i)|=\vspace{0.3cm}\\
\displaystyle   =
|{\cal G}(m,0)|+|{\cal G}(m,1)|+\ldots+|{\cal G}(m,k)|,
\end{array}
$$

\noindent where $|{\cal G}(m,i)|$ is either the cardinality of the new obtained PN with $k-i$ tokens in the added place or that one of the initial PN with $i$ tokens.
Such results can be reported in a matrix form, obtaining the well known Pascal matrix, that comes out from the Pascal's triangle
.
The elements of the symmetric Pascal matrix are the binomial coefficients, i.e., it holds that
$$\binom{i+j-2}{i-1}$$
having $ i=m$, $j=k+1$. \hfill $\square$}

Considering the above result, one can state the following lemma.

\begin{lemma} \label{lemma:classicONsmPN}\rm
Consider a strongly connected SM with $k$ tokens. Let $m$ be the number of its places and $E$ be its input alphabet.
For such a net, Algorithm~\ref{algo_ss_PN} requires a time
$$
\begin{array}{l}
\hspace{1.5cm} \displaystyle   O(|{\cal G}(m,k)|^3+|E||{\cal G}(m,k)|^2) \vspace{0.3cm}\\
\hspace{1cm} \displaystyle   \leq
O\left(\left[\frac{k^{m-1}}{(m-1)!}\right]^3
+|E|\left[\frac{k^{m-1}}{(m-1)!}\right]^2\right).\qquad\;\hspace{1cm} \blacksquare
\end{array}$$

\end{lemma}

\subsection{Complexity via synchronizing transition sequences}

We have shown in Proposition~\ref{propo_1token} a technique to compute a $1$-SS on a strongly connected net based on synchronizing transitions sequences. Here we discuss the complexity of such a procedure.

To compute a synchronizing transition sequence one can proceed using a backward depth-first search from place $\bar{p}$ and verifying the conditions of Definition~\ref{synch_path} over the labeling function. 


%
It is known that a depth first search requires $O(b^d)$ time \cite{algorithms01}, for explicit graphs traversed with repetition, having a branching factor $b$ and a depth search of $d$.

Assume that a SM has a backward branching factor (the number of transitions inputting in a place) bounded by $\phi = \max_{p \in P} |^{\bullet} p|$. While exploring the net with possible repetitions of places, an upper bound for the depth search length is $q-1$, where $q$ is the number of net transitions. Thus a first very rough approximation of the needed time is given by $O(\phi^{q-1})$.

This time only depends on structural net parameters, does not grow with the number of tokens and is typically smaller than the time required by Algorithm~\ref{algo_ss_PN}.

%
%
%
%
%
%

\section{Synchronizing sequences on non-strongly connected state machines}\label{NSCPN}

Consider now connected --- but not necessarily strongly connected --- state machines. It can be shown how the existence of a SS depends on the interconnection between ergodic and transient components.

\begin{proposition}\label{NO_SS_onPN}\em
Consider a synchronized SM $\langle N, E,f\rangle$ with $\mu$ transient components and $\eta$ ergodic components.
If $\eta>1$ there exists no SS for such a net.\end{proposition}
\IEEEproof{Let the net have two ergodic components  $ER'$ and $ER''$. Consider two initial markings $M'_{0}$ and $M''_{0}$ both with $k$ tokens
such that $M'_0$ (resp., $M''_0$) assigns all tokens to the component $ER'$ (resp., $ER''$). Clearly there exists no marking $\bar{M}$ reachable from both $M'_0$ and $M''_0$, hence no SS exists according to Definition~\ref{def_SS1}.\hfill $\square$}

It is now proposed an algorithm to determine sequences for not strongly connected state machines having a single ergodic component where the interconnection between transient components can be arbitrary.


It is first stated the following result.
\begin{proposition}\label{struc_A_N}
\em
Consider a SM $N=(P,T,Pre,Post)$ with a single component ER and let ${\cal C}(N)$ be its condensed graph. For each node $v_i$ of ${\cal C}(N)$ associated with a transient component $TR_i$ (with $i > 0$), let $l_i$  be the length of the longest path from $v_i$ to node $v_0$ associated with the ergodic component $ER$. Then if there is an edge $(v_i, v_j) $ in ${\cal C}(N)$ it holds $l_i > l_j$.
\end{proposition}

\IEEEproof{First observe that ${\cal C}(N)$ is acyclic by construction and the node $v_0$ is reachable from any other node, hence $l_j \in \nat$ is well defined for each node $v_j$ (with $j > 0$).
By definition, if  $(v_i, v_j) $ is an edge of  ${\cal C}(N)$, then  $l_i \geq l_j +1$. \hfill $\square$}

The following algorithm for the one-token case allows to obtain a SS, such that a place $\bar{p}$ in the single ergodic component is marked.

\begin{algo}\textbf{(Computing a SS leading to $\bar{p}\in ER$)}
\label{algo_ss_1ER}
\rm

\noindent\textbf{Input:} A synchronized PN $\langle N,E,f\rangle$ containing $1$ tokens, with $\mu$ transient components and $1$ ergodic component.
\\\textbf{Ouput:} A SS $\bar{w}$ for place $\bar{p}$.
\begin{enumerate}
\item [{\bf 1.}] Let ${\cal C}(N)$ be the condensed graph of $N$ and associate ER with node $v_{0}$.
\item [{\bf 2.}] Label every other node $v_i$ of ${\cal C}(N)$ with $l_i$, where $l_i$ is the length of the longest path from $v_i$ to $v_0$.
\item [{\bf 3.}] Let $\Sigma_k$ be the set of nodes such that $\Sigma_k=\{v_i: l_i=k\}$, thus for construction $\Sigma_0=\{v_0\}$.
\item [{\bf 4.}] Let $w=\varepsilon$.
\item [{\bf 5.}] For k=$l_{max}$:1
\begin{enumerate}
\item [{\bf 5.1.}] for all $v_i \in \Sigma_k$,
\begin{enumerate}
\item [{\bf 5.1.1.}] pick any transition $t'$ connecting $v_i$ to $v_j$, being $v_j \in \Sigma_{k'}$ and $k'< k$;
\item [{\bf 5.1.2.}] consider the strongly connected subnet associated with node $v_i$. Determine a $1$-SS $w'$ for place $p'$, where $p'\in t'^\bullet$;

\end{enumerate}
\end{enumerate}
\item [{\bf 6.}] Consider the strongly connected subnet associated with node $v_0$. Let $w$ be a $1$-SS for place $\bar{p}$.

\item [{\bf 7.}] Let $w=wf^*(\sigma)$. \hfill $\blacksquare$
\end{enumerate}
\end{algo}

The algorithm starts taking into account the farthest nodes from ER. By definition of condensed graph, transient nodes with the same label value are not connected. Hence at step~5.1.2. the application of each couple $(w', t')$, step by step, drives the token always nearer to ER until it reaches it.

We have remarked that a net with more than one ergodic component cannot have a SS, at least according to Proposition~\ref{NO_SS_onPN}. However, the knowledge of the initial token distribution among the net components may lead to other interesting characterizations, provided of course the initial state uncertainty is redefined according to this new information.

{

\section{Synchronizing sequence on nets with state machine subnets}\label{sec:subSMs}
In the following we discuss some results on synchronized PNs which do not belong to the class of SMs and show how --- under certain conditions --- our approach can still be applied in this more general setting.



\begin{proposition}
Consider a bounded synchronized PN $\langle N, E,f \rangle$. Let $P=P_s\cup P_z$ and $T=T_s\cup T_z$,
such that $N_s=\left(P_s,T_s,Pre_s,Post_s \right)$ is a strongly connected SM subnet, where $Pre_{s}$ and $Post_{s}$ are the restrictions of $Pre$ and $Post$ to $P_{s}\times T_{s}$.

Let $\bar{w}$ be a SS that drives the subnet $N_s$ to a target marking $\bar{M}_s$. This sequence $\bar{w}$ drives $N$ to a target marking $\bar{M}$ such that:
$$\bar{M}(p) =\bar{M}_s(p) \;\;if\;\; p\in P_s,$$
\noindent if the two following conditions hold:
\begin{description}
\item[i)] $\{T_z^\bullet\cup^\bullet T_z\}\cap P_s=\emptyset$;
\item[ii)] ($\forall e\in \bar{w}$) $T_e\cap P_z^\bullet\cap T_s=\emptyset$.
\end{description}

\IEEEproof{
Condition i) states that no transition $t\in T_z$ is connected to any place $p\in P_s$. This ensures that the firing of a transition in $T_z$ cannot affect the marking of places in $P_s$. Hence, given the special structure of $N_s$, the following condition holds for any initial marking $M_0$:
${\displaystyle\left(\forall M\in R(N,M_0)\right)\;\;\sum_{p\in P_s}M(p) = \sum_{p\in P_s}M_0(p),}$
i.e., the token count in the SM component remains constant.


\begin{table*}[htbp]
\centering
\subtable[$|N_{STS}|/|N_{RG}|$ \label{tab:N_1}]{%
\begin{tabular}{|c|c|c|c|c|c|c|c|c|c|c|c|c|c|}
\hline \backslashbox{$m$}{$q$} &$3$	   &$4$	&$5$     &$6$	 &$7$      &$8$&$9$	&$10$&$11$&$12$&$13$&$14$&$15$ \\
\hline$2$&$1$	   &$1$	&$1$     &$1$	 &$1$      &$1$&$1$	&$1$&$1$&$1$&$1$&$1$&$1$\\
\hline$3$&$1$&$1$	&$1$     &$1$	 &$1$      &$1$&$1$	&$1$&$1$&$1$&$1$&$1$&$1$\\
\hline$4$&\cellcolor{black}&$1$&$1$	     &$1$	 &$1$      &$1$&$1$	&$1$&$1$&$1$&$1$&$1$&$1$\\
\hline$5$&\cellcolor{black}&\cellcolor{black}&$1$&$0,8$	 &$1$      &$1$&$1$	&$1$&$1$&$1$&$1$&$0,8$&$1$\\
\hline$6$&\cellcolor{black}&\cellcolor{black}&\cellcolor{black}&$1$&$1$     &$1$&$1$	&$0,8$&$1$&$1$&$1$&$1$&$1$\\
\hline$7$&\cellcolor{black}&\cellcolor{black}&\cellcolor{black}&\cellcolor{black}&$1$&$1$&$0,6$&$0,66$&$0,8$&$1$&$0,88$&$1$&$1$\\
\hline
\end{tabular}
}\qquad\qquad

\subtable[$|\hat{T}_{STS}\,|/|\hat{T}_{RG}|$\label{tab:T_1}]{%
\begin{tabular}{|c|c|c|c|c|c|c|c|c|c|c|c|c|c|}
\hline \backslashbox{$m$}{$q$}&$3$&$4$&$5$&$6$&$7$&$8$&$9$&$10$&$11$&$12$&$13$&$14$&$15$\\
\hline$ 2$ & $0.22$ &$0.14$ &$0.24$ &$0.17$ &$0.20$ &$0.15$ &$0.13$ &$0.18$ &$0.15$ &$0.18$ &$0.12$ &$0.12$ &$0.10$ \\
\hline$ 3$ & $0.23$ &$0.23$ &$0.21$ &$0.26$ &$0.23$ &$0.26$ &$0.36$ &$0.48$ &$0.54$ &$0.30$ &$0.33$ &$0.40$ &$0.74$ \\
\hline$ 4$ & \cellcolor{black} &$0.43$ &$0.20$ &$0.32$ &$0.37$ &$0.50$ &$0.62$ &$0.54$ &$0.52$ &$0.65$ &$0.61$ &$0.75$ &$0.93$ \\
\hline$ 5$ & \cellcolor{black} &\cellcolor{black} &$0.6$ &$0.30$ &$0.32$ &$0.41$ &$0.53$ &$0.60$ &$0.90$ &$0.59$ &$0.78$ &$0.72$ &$1.52$ \\
\hline$ 6$ & \cellcolor{black} &\cellcolor{black} &\cellcolor{black} &$0.45$ &$0.21$ &$0.29$ &$0.45$ &$0.72$ &$0.73$ &$0.54$ &$1.03$ &$0.76$ &$1.08$ \\
\hline$ 7$ & \cellcolor{black} &\cellcolor{black} &\cellcolor{black} &\cellcolor{black} &$0.62$ &$0.16$ &$0.22$ &$0.27$ &$0.88$ &$0.78$ &$1,15$ &$1,6$ &$2,54$\\

\hline
\end{tabular}
}

\subtable[$|\hat{L}_{STS}\,|$/$|\hat{L}_{RG}|$\label{tab:L_1}]{%
\begin{tabular}{|c|c|c|c|c|c|c|c|c|c|c|c|c|c|}
\hline \backslashbox{$m$}{$q$}&$3$&$4$&$5$&$6$&$7$&$8$&$9$&$10$&$11$&$12$&$13$&$14$&$15$\\
\hline$ 2$ & $1.00$ &$1.00$ &$1.00$ &$1.00$ &$1.00$ &$1.00$ &$1.00$ &$1.00$ &$1.00$ &$1.00$ &$1.00$ &$1.00$ &$1.00$ \\
\hline$ 3$ &$0.83$ &$0.91$ &$0.91$ &$0.91$ &$0.83$ &$1.00$ &$1.10$ &$0.83$ &$1.00$ &$0.91$ &$0.77$ &$1.10$ &$1.00$ \\
\hline$ 4$ & \cellcolor{black} &$0.65$ &$0.85$ &$0.89$ &$1.00$ &$0.77$ &$0.85$ &$0.95$ &$0.84$ &$1.00$ &$0.94$ &$0.89$ &$1.06$ \\
\hline$ 5$ & \cellcolor{black} &\cellcolor{black} &$0.8$ &$0.85$ &$0.85$ &$0.70$ &$0.91$ &$0.91$ &$0.96$ &$0.75$ &$1.00$ &$0.87$ &$0.92$ \\
\hline$ 6$ & \cellcolor{black} &\cellcolor{black} &\cellcolor{black} &$0.6$ &$0.53$ &$0.90$ &$0.79$ &$0.90$ &$0.80$ &$0.74$ &$0.86$ &$0.81$ &$0.70$ \\
\hline$ 7$ & \cellcolor{black} &\cellcolor{black} &\cellcolor{black} &\cellcolor{black} &$0.54$&$0.61$ &$0.57$ &$0.50$ &$0.88$ &$0.59$ &$0,76$ &$0,85$ &$0,69$ \\
\hline
\end{tabular}
}
\caption{Numerical results for randomly generated SM PNs ($k = 1$)}\label{tab:1tok}
\end{table*}

\begin{table*}[htbp]
\centering

\subtable[$|\hat{T}_{STS}\,|/|\hat{T}_{RG}|$\label{tab:T_2}]{%
\begin{tabular}{|c|c|c|c|c|c|c|c|c|c|c|c|c|c|}
\hline \backslashbox{$m$}{$q$}&$3$&$4$&$5$&$6$&$7$&$8$&$9$&$10$&$11$&$12$&$13$&$14$&$15$\\
\hline$ 2$ & $0.26$ &$0.12$ &$0.23$ &$0.16$ &$0.17$ &$0.13$ &$0.11$ &$0.16$ &$0.14$ &$0.15$ &$0.11$ &$0.11$ &$0.10$ \\
\hline$ 3$ & $0.27$ &$0.16$ &$0.14$ &$0.14$ &$0.14$ &$0.21$ &$0.28$ &$0.35$ &$0.42$ &$0.24$ &$0.27$ &$0.24$ &$0.54$ \\
\hline$ 4$ & \cellcolor{black} &$0.28$&$0.12$ &$0.20$ &$0.23$ &$0.31$ &$0.41$ &$0.35$ &$0.11$ &$0.39$ &$0.20$ &$0.36$ &$0.40$ \\
\hline$ 5$ & \cellcolor{black} &\cellcolor{black} &$0.17$ &$0.08$ &$0.08$ &$0.19$ &$0.01$ &$0.28$ &$0.44$ &$0.27$ &$0.35$ &$10^{-3}$ &$0.75$ \\
\hline$ 6$ & \cellcolor{black} &\cellcolor{black} &\cellcolor{black} &$10^{-3}$ &$10^{-3}$ &$0.09$ &$0.02$ &$10^{-3}$ &$10^{-3}$ &$0.20$ &$10^{-3}$ &$0.29$ &$0.39$ \\
\hline$ 7$ & \cellcolor{black} &\cellcolor{black} &\cellcolor{black} &\cellcolor{black} &$10^{-3}$&$10^{-3}$ &$10^{-3}$ &$0.05$ &$0.05$ &$10^{-3}$ &$10^{-3}$ &$10^{-4}$ &$10^{-3}$ \\
\hline
\end{tabular}
}

\subtable[$|\hat{L}_{STS}\,|$/$|\hat{L}_{RG}|$\label{tab:L_2}]{%
\begin{tabular}{|c|c|c|c|c|c|c|c|c|c|c|c|c|c|}
\hline \backslashbox{$m$}{$q$}&$3$&$4$&$5$&$6$&$7$&$8$&$9$&$10$&$11$&$12$&$13$&$14$&$15$\\
\hline$ 2$ & $1.71$ &$1.71$ &$1.71$ &$1.71$ &$1.71$ &$1.71$ &$1.71$ &$1.71$ &$1.71$ &$1.71$ &$1.71$ &$1.71$ &$1.71$ \\
\hline$ 3$ &$1.5$ &$1.60$ &$1.41$ &$1.60$ &$1.33$ &$1.71$ &$1.89$ &$1.50$ &$1.71$ &$1.60$ &$1.41$ &$1.89$ &$1.76$ \\
\hline$ 4$ & \cellcolor{black} &$1.5$ &$1.41$ &$1.42$ &$1.64$ &$1.46$ &$1.57$ &$1.60$ &$1.54$ &$1.54$ &$1.50$ &$1.54$ &$1.77$ \\
\hline$ 5$ & \cellcolor{black} &\cellcolor{black} &$1.1$ &$1.42$ &$1.35$ &$1.26$ &$1.44$ &$1.50$ &$1.55$ &$1.36$ &$1.65$ &$1.37$ &$1.65$ \\
\hline$ 6$ & \cellcolor{black} &\cellcolor{black} &\cellcolor{black} &$1.3$ &$0.92$ &$1.52$ &$1.32$ &$1.64$ &$1.27$ &$1.33$ &$1.28$ &$1.39$ &$1.30$ \\
\hline$ 7$ & \cellcolor{black} &\cellcolor{black} &\cellcolor{black} &\cellcolor{black} &$0.85$&$0.97$ &$0.97$ &$0.92$& $0.88$ &$0.77$ &$0,79$ &$0,64$ &$0,79$ \\
\hline
\end{tabular}
}
\caption{Numerical results for randomly generated SM PNs ($k = 2$)}\label{tab:2tok}
\end{table*}

Let $\bar w=e_1 e_2 \cdots e_k$ be a SS for subnet $N_s$ that yields a known marking $\bar{M}_s$ from any reachable marking $M$ of the subnet. To prove the result, it is sufficient to show that at each step $i=1, \ldots, k$ the same sequence, applied to $N$ from any marking $M'$, with $M'(p) =M(p)$ if $p \in P_s$, produces exactly the same transition firings that it produces in $N_s$.

\begin{figure}[t]
 \centering
 \subfigure[]
   {\includegraphics[scale=0.6]{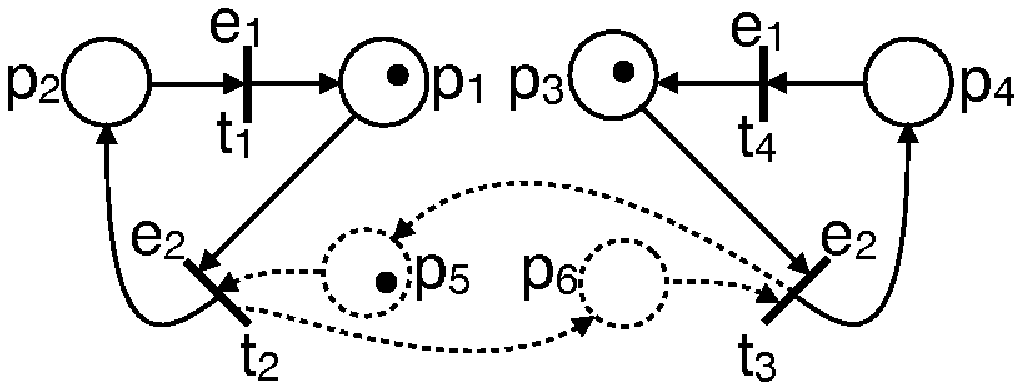}\label{fig:GMEC2}}\\
 \subfigure[]
   {\includegraphics[scale=0.9]{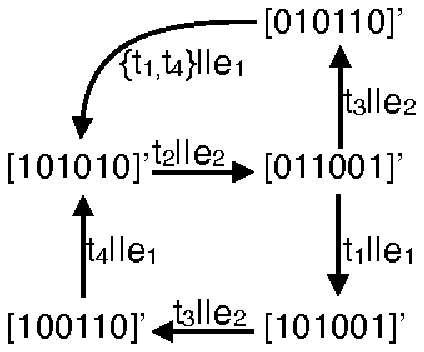}\label{fig:GMEC2_RG}}\hspace{0,5cm}
\caption{A synchronized PN (a) and its RG}
\label{fig:gmec}
 \end{figure}

In fact, when a input symbol $e_i \in \bar w$ is applied:
\begin{itemize}
\item all transitions that can fire in $N_s$ can also fire in $N$, because the additional places $P_z$ in $N$ cannot disable these transitions since they do not belong to $^\bullet T_s$;
\item no transition in $P_z^\bullet$ can fire, because no transition in $P_z^\bullet$ has label $e_i$.\hfill $\square$
\end{itemize}
}\label{propo:notSM}
\end{proposition}

Such a result can be further generalized to nets containing more than one state machine subnets.

\begin{proposition}
Consider a bounded synchronized PN $\langle N, E,f \rangle$. Let $P_s\cup P_z=P$ and $T_s\cup T_z=T$, where $\displaystyle P_s=\bigcupdot_{i=1}^nP_{s,i}$ and $\displaystyle T_s=\bigcupdot_{i=1}^nT_{s,i}$ (here $\bigcupdot$ denotes the union of disjoint subsets).
These sets are such that for $i=1,2,\dots n$ $N_{s,i}=\left( P_{s,i},T_{s,i},Pre_{s,i},Post_{s,i} \right)$ is a strongly connected SM subnet. $Pre_{s,i}$ and $Post_{s,i}$ are the restrictions to $Pre$ and $Post$ to $P_{s,i}\times T_{s,i}$.

For every subnet $N_{s,i}$, let $\bar{w}_i$ be a SS that drives the subnet $N_{s,i}$ to a target marking $\bar{M}_{s,i}$.
The sequence $\bar{w}=\bar{w}_1\bar{w}_2\ldots\bar{w}_n$ drives $N$ to a target marking $\bar{M}$ such that:
$$\bar{M}(p)=\bar{M}'_{s,i} \; with \; \bar{M}_{s,i}\xrightarrow{\bar{w}_{i+1}\bar{w}_{i+2}\cdots\bar{w}_n}\bar{M}'_{s,i}\;\;\;\;if\;\; p\in P_{s,i},$$

\noindent if the two following conditions hold:
\begin{description}
\item[i)] $\{^\bullet T_z\cup T_z^\bullet \}\cap P_s=\emptyset$;
\item[ii)] ($\forall e\in \bar{w_i}$) $\displaystyle T_e\cap P_z^\bullet\bigcap_{j=1}^iT_{s,j}=\emptyset$.
\end{description}

\IEEEproof{The proof follows along the same lines of the proof of Proposition~\ref{propo:notSM} with just an additional consideration.
First we observe that condition ii) in Proposition~\ref{propo:notSM_plus}  is a generalization of condition ii) in Proposition~\ref{propo:notSM}: the condition now must hold not only for the transitions of net $i$ but also for those of all nets $j$ with $j <i$. In fact, the overall SS is composed by concatenation of the SSs for each state machine subnet. When we apply the SS $w_i$ to the net, we assume that the markings of all subnets $j$, for $j < i$ are known but may change, as some transitions in the already synchronized subnets may be receptive to an event $e \in \bar w_i$. However, condition ii) ensures that the enabling condition of these transitions does not depend on the places  in $P_z$, whose marking is unknown, and the marking reached after the application of event $e$ is computable.}
\hfill $\square$\label{propo:notSM_plus}
\end{proposition}

Sequence $\bar w$ determined in the previous proposition is a SS for the subnet $N_s$. It also drives the complete model $N$ to a state where the marking of places in $P_s$ is known, while in general nothing can be said about the marking of places in $P_z$.


\begin{example}
Consider the net in Figure~\ref{fig:GMEC2}. Let $P_{s,1}=\{p_1,p_2\}$, $P_{s,2}=\{p_3,p_4\}$, $P_z=\{p_5,p_6\}$, $T_{s,1}=\{t_1,t_2\}$, $T_{s,2}=\{t_3,t_4\}$ and then $T_z=\emptyset$. $N_s$ is then the net depicted in Figure~\ref{fig:GMEC2}, without taking into account dashed places and arcs.
Let $\bar{w_1}$ and $\bar{w_2}$ be SSs that drives respectively $N_{s,1}=\left(P_{s,1},T_{s,1},Pre_{s,1},Post_{s,1}\right)$ to $\bar{M}_{s,1}=[0\,1]^T$ and $N_{s,2}=\left(P_{s,2},T_{s,2},Pre_{s,2},Post_{s,2}\right)$ to $\bar{M}_{s,2}=[0\,1]^T$.
By separately analyzing the two subnets, $\bar{w}_1=\bar{w}_2=e_1$ are obtained.
$\bar{w}=\bar{w}_1\bar{w}_2$ respects conditions i) and ii) of Proposition~\ref{propo:notSM_plus} and is therefore a SS for $N_s$, i.e., it drives the net to a marking $\bar{M}$ that is either $[1\,0\,1\,0\,1\,0]^T$ or $[1\,0\,1\,0\,0\,1]^T$, as can be seen by its RG in Figure~\ref{fig:GMEC2_RG}.
\hfill $\square$
\end{example}


\section{Experimental results}\label{sec:ex_results}

This section has two objectives. First, we compare the two algorithms we have presented for SS computation of state machine Petri nets (reachability based versus path based) by applying them to  randomly generated nets and analyzing their performance. The model data and MATLAB programs can be downloaded from \cite{Pocciweb}.

All simulations have been run on a mini Mac intel core Duo $2$, $2.53$ GHz processor, with $4$ GB $1067$ MhZ DDR3 RAM.




Randomly generated models have been previously adopted as a validation method for synchronizing sequence construction also by Roman \cite{Roman2009125}.

For selected values of $m$ places, $q$ transitions and $k=1,2$ tokens, we randomly generate $100$ deterministic and strongly connected synchronized SMs having $m=2 \div 7$ places, $q = m \div 15$ transitions and $k=1,2$ tokens. In all cases the input alphabet has cardinality $f$ randomly chosen in $\frac{q}{m} \div q$. Note that $\frac{q}{m}$ is the minimal alphabet cardinality to ensure the determinism for a SM having $m$ places and $q$ transitions.
For each net a place is randomly selected and we use both Algorithm~\ref{algo_ss_PN} (denoted RG) and Algorithm~\ref{algo_ss_PN} (denoted by STS) to determine a SS to this place. The algorithms are compared by means of three performance indexes:
\begin{description}
\item[$N_{RG}$, $N_{STS}$:]\hspace{1cm}number of times the algorithm successfully terminates returning a SS;
\item[$\hat{T}_{RG}$, $\hat{T}_{STS}$:]\hspace{1cm}average time required to compute the sequence;
\item[$\hat{L}_{RG}$, $\hat{L}_{STS}$:]\hspace{1cm}average length of the sequences.
\end{description}

Finally the performance of the two approaches is evaluated by computing the ratio of $N_{STS}$ (resp. $\hat{T}_{STS}$ and $\hat{L}_{STS}$) to $N_{RG}$ (resp. $\hat{T}_{RG}$ and $\hat{L}_{RG}$).

Results are shown in Table~\ref{tab:1tok} for nets with one token and in Table~\ref{tab:2tok} for the two token case. Note that the table showing $N_{STS}$/$N_{RG}$ does not depend on the number of tokens and thus it is shown only for $k=1$. Black cells denote parameter values for which no strongly connected SM can be generated, i.e., for $m>q$.


Table~I(a) shows the ratio $N_{STS}$/$N_{RG}$ between the number of times a SS has been found using the STS and the RG approach. In the previous sections we have mentioned that while the RG approach always determines a SS if any exists, the STS approach may fail to do so. Hence the value in the table should be contained in the interval $[0, 1]$. We can observe, however, that over 88\% of the table entries show a value of $1$, hence confirming that the STS approach can find a solution in most cases and thus this result is not too restrictive.

Table~I(b) shows the ratio $T_{STS}$/$T_{RG}$ between the execution time to compute a SS using the STS and the RG approach for nets with one token. Here we expect the STS approach to be more efficient, as discussed in Section~\ref{complexity}, and this is confirmed from the fact that in almost all cases the table entries are smaller than one. Only in a few cases, for very large values of $m$ and $q$, we have that the RG method is faster than the STS one. This, we believe, it is due to our implementation of the STS approach that uses a brute force depth-first search.

\begin{figure}[t]
\centering
  \includegraphics[scale=0.6]{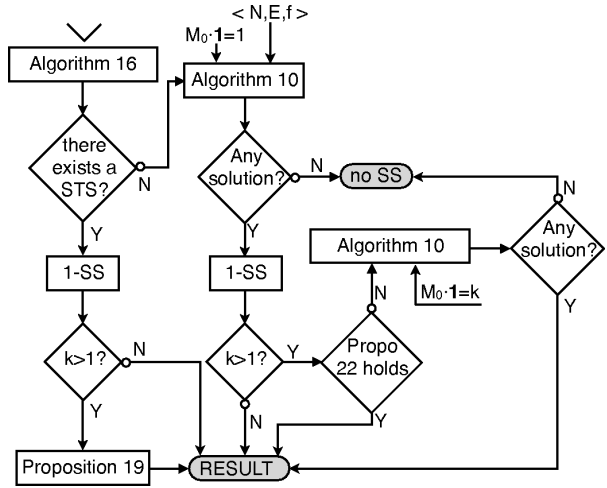}\\
  \caption{Suggested way to determine a SS for strongly connected SMs.}\label{fig:flowchart}
\end{figure}

Table~II(a) shows the ratio $T_{STS}$/$T_{RG}$ between the execution time to compute a SS using the STS and the RG approach for nets with two tokens. Here we see the main advantage of the STS method, that can use a $1$-SS to determine a $k$-SS, while the RG method had a complexity that grows polynomially with $k$ (and exponentially with $m$), as discussed in Section~\ref{complexity}. Here the advantage of the STS method is more noticeable for large values of $m$ and $q$.
Table~I(c) and Table~II(b) show the ratio $L_{STS}$/$L_{RG}$ between the average length of a SS computed using the STS and the RG approach for nets with one or two tokens. This index is probably less significant than the previous ones, although one may argue that the shortest the SS the less expensive is the synchronization (in terms of costs or of time required). Here we can see that in the case of one token the STS approach in most of the cases produces shorter SS, while the situation is the opposite for 2 tokens. This is due to the fact that the $k$-SS obtained by STS is always $k$ times longer than the corresponding $1$-SS, while shorter solutions may be obtained by the RG approach.

On the base of these results, we can say that to compute a SS for strongly connected  Petri nets it is convenient to first search for a STS based solution using Algorithm~\ref{algo_ss_STS} and then, if this fails, to use Algorithm~\ref{algo_ss_PN}.
 This is summarized in the flowchart in Figure~\ref{fig:flowchart}.}

\section{Conclusions and future works}\label{conclusion}

In this paper, we have shown how automata techniques can be applied with minor changes to the class of bounded synchronized PNs.

Also it has been proposed a method that allows to determine a Synchronizing Sequence for the class of synchronized state machine PNs.


Our approach alleviates the state explosion problem also in the case of multiple tokens, since the construction of the reachability graph is not needed. 
We have shown by means of several examples
how the computational time does not increase as the number of tokens in the net increases

{There is an open line for interesting future works. We plan to extend our approach to unbounded PNs, whose behavior can be approximated with a finite \emph{coverability graph} (CG), by introducing an $\omega$ component to denote a place whose token content may be arbitrarily large.


Note that classic coverability methods construction cannot be directly applied to this class of PNs, 
that is why a new algorithmic procedure for the CG construction has to be provided.


}

\bibliography{biblio.bib}

\end{document}